\documentstyle[prb,aps]{revtex} 

\begin{document}

\title{Reflectivity Anisotropy Spectra  of Cu- and Ag- (110) surfaces from {\it ab initio} theory   }

\author{ Patrizia Monachesi$^*$, Maurizia Palummo, Rodolfo Del Sole } 

\address{Dipartimento di Fisica - Universita' di Roma, 'Tor Vergata' and {\it Istituto Nazionale di Fisica della Materia}
I-00133 Roma, Italy}

\author{Rajeev Ahuja and Olle Eriksson}

\address{Condensed Matter Theory Group,Physics Department, Uppsala University, S-7512 Uppsala, Sweden}
\maketitle
 
\begin{abstract}

We present the surface Reflectivity Anisotropy spectra of the relaxed (1x1)  (110) surface  of Cu and Ag as obtained 
 by {\it ab initio} calculations. 
We are able to disentagle the effects of the intraband and interband parts of the bulk  dielectric 
function on the bare dielectric  anisotropy of the surface. 
We show  how  the position,  sign and   amplitude of the  structures observed in such
spectra depend on the above quantities.  The  lineshape of all the calculated structures   agree very well with 
the ones observed experimentally for samples treated by suitable surface cleaning. In particular, we reproduce the observed 
single peak structure of Ag at high energy, found to represent a   state of the clean surface different from the one giving the
originally observed double peak structure. This results is not reproduced by the 'local field' model. 

%The  interband part of  surface and bulk dielectric  functions  
%are computed by   self-consistent electronic structure calculations with the FP-LMTO method within LDA approximation. 
%The intraband part of the bulk dielectric function, also known as Drude tail, is computed using the plasma frequency estimated from
%experiments. 
  
\end{abstract}  

The Reflectivity  Anisotropy Spectroscopy (RAS) is a powerful technique to probe the optical properties of an anisotropic  medium 
like the surface  of an otherwise isotropic  bulk crystal. 
The RAS  is measured as the relative difference $\frac{\Delta r} {r}$ of the complex reflectivity $r$ along two perpendicular 
axes on the surface.  Despite its vast application in semiconductors 
surface science since the pioneering work by Aspnes\cite{aspnes}, the first RAS spectrum of a metallic surface appeared as late as 
in 1993 for the Ag (110) surface\cite{agras}. The characteristic 
 resonance-like structure at the bulk plasma frequency, i.e. at $\omega_p$ = $\simeq$ 3.8 eV, has been considered a fingerprint of
the Ag-(110) surface. More recently, the research on the optical properties of metallic surfaces has gained 
renewed impulse and in the  last three years the (110) surface of Ag and Cu have been investigated  both by optical 
 and photoemission spectroscopy\cite{stahr98,fernandez97,hansen98,bremer99,stahr00}. 
 
The RAS results essentially agree in establishing the existence of one peak 
at low energy ($<$ 2.5 eV) of similar shape  in both Cu and Ag and of a further structure 
at higher energy  of quite different shape.   In particular, some authors point out  that the lineshape 
of the high-energy peaks  depends critically upon the surface treatment during  the sample preparation\cite{hansen98,bremer99}. 
This experimental aspect is very important  for a correct comparison  of RAS spectra with theoretical models and calculations. 

A major problem with the interpretation of the RAS spectra is that  the optical anisotropy  introduced  by the  surface
is   difficult to extract from  
the measured reflectivities  without a solid quantitative theoretical basis.   An attempt to do so
 has been made from a phenomenological point of view\cite{cole98}.  Experimentally one can   observe  the evolution of a given 
 feature  in the spectra from the clean 
 to the gas-covered  surface. In fact coverage by, e.g. oxygen, is known to wash out progressively optical structures  originating 
 from surface-to-surface state transitions, as a function of time and/or amount of oxygen.

  RAS structures observed in the noble metals are currently interpreted in terms of interband 
transitions between  surface or bulk states by speculating on available 
results of existing band structure obtained by folding bulk bands or by interpolation methods. Another tool of  
interpretation is the  so-called 'swiss cheese' model\cite{mochan94}, describing the optical response of the surface by layer-depending 
dipolar interactions.   
This phenomenological  model partially   succeeds, at least qualitatively,   to explain the RAS  peaks observed in Ag and Cu under suitable 
experimental conditions\cite{hansen98,bremer99}. The lineshape, however, depends on the parameters of the model\cite{hansen98}.
Only very recently first principle calculations of the optical properties of noble metals surfaces have started to 
appear\cite{mrs99,ecoss99-bouarab}, opening the possibility of a direct quantitative interpretation of the measured optical spectra.  
Such self consistent calculations yield, on the same footing, the  electronic  structure and  the dielectric 
functions  $\epsilon_{[1\overline{1}0]}$,  $\epsilon_{[001]}$ along the directions $[1\overline{1}0]$ and $[001]$  of the (110) surface 
. The  dielectric function is computed from interband transitions among electronic states. From these, 
$\Delta\epsilon$= $\epsilon_{[1\overline{1}0]}-\epsilon_{[001]}$, known as Surface Dielectric Anisotropy (SDA), is derived.
So far,  comparison of the calculated SDA with direct experimental RAS results is  made 
 through a reliable semi-phenomenological model\cite{cole98,sda} that transforms the RAS spectrum into the SDA 
 via the bulk dielectric function $\epsilon^b$ normally obtained by independent measurements. 
 Attempts to interpret the RAS spectrum in terms of  bulk or surface  electronic transitions from the SDA structures
  may, however,  be incorrect\cite{ecoss99-bouarab}. In fact,  the passage from the anisotropic
 dielectric function  to the RAS spectrum involves bulk quantities that re-shape much  of the  bare  metallic SDA spectrum. 
 At low energies, as shown below,  RAS and SDA have different signs due to intraband transitions, at variance with the case of
  semiconductors. This point is even more important in view of recent  controversial  results for the sign of the 
  RAS\cite{stahr00,hofmann95}.  
   Therefore,  as we aim to show in this Letter,
 only the  calculation of the RAS spectrum from consistently computed quantities allows to distinguish between RAS structures arising 
 from the bare SDA from those related to bulk effects.   
\vskip 1cm

We shortly summarize the fundamental analytical expressions necessary for the  analysis of a RAS spectrum. 
 The surface reflectivity anisotropy $\frac{\Delta r} {r}$,  written for the present  surface geometry,  is 
$ \frac{\Delta r} {r} = 2 \frac{r_{[1\overline{1}0]}-r_{[001]}} {r_{[1\overline{1}0]}+r_{[001]}}  $. 
The RAS spectrum , as obtained directly from reflectivity measurements,  is usually given as the real part of the above complex 
quantity, namely:

$$Re\{ \frac{\Delta r} {r}\} = C (\frac{\Delta\epsilon_2(\epsilon^b_1-1)}{(\epsilon^b_1-1)^2+(\epsilon^b_2)^2} - \frac{\epsilon^b_2
  \Delta\epsilon_1} {(\epsilon^b_1-1)^2+(\epsilon^b_2)^2}),$$

where subscripts $1,2$ refer to the real and imaginary parts, respectively and C is a constant that is of no importance here.

In metals, all the $\epsilon$'s in the formula above  contain  terms due to interband as well intraband transitions.
The latter ones give rise to the so-called Drude tail dominating at  low energies.
We  assume that  the anisotropy of the intraband part of $\Delta\epsilon$ is negligible, which allows us to neglect
intraband effects {\it tout court} in this term. In fact  there is, so far, 
some  experimental evidence\cite{bremer99} that the intraband anisotropy of the surface affects the spectrum of Cu only up to 1.5 eV.
The inclusion of intraband effects in  $\epsilon^b$ is, instead, of paramount importance, as we will show below.
Therefore our calculations  contain the {\it ab initio} computation of the interband part of all the
surface and bulk $\epsilon$'s plus the inclusion of  the Drude tail in $\epsilon^b$ computed by the standard relations involving the 
plasma frequency $\omega_p$ and the relaxation time $\tau$ of the bulk metal\cite{johnson&christy}.

The computation of {\it ab initio} quantities, i.e., electronic properties and intraband part of the dielectric functions, has been 
carried out self-consistenly by the  Full Potential version\cite{wills} of the Linear Muffin Tin Orbital method\cite{andersen} 
within the Local Density Approximation\cite{kohn-sham} for  the exchange and correlation potential\cite{XC}. 
The  surface geometry is obtained within a  repeated slab scheme including 11 atomic plus 6 vacuum layers. 
The reciprocal space integrations  are performed with the analytical tetrahedron method\cite{tetrah}. Meshes  as dense as 
752 and 256 k-points in the
irreducible bulk and surface Brillouin Zone, respectively, give converged values for the dielectric functions calculated as 
matrix elements of the momentum operator among occupied and unoccupied electronic states. The relaxed\cite{unrelx}, unreconstructed
  geometry of the Cu and Ag (110) surfaces\cite{relx} has been taken into account.
A more complete description of the  details of the calculations will be published in a future work.

In Fig.\ \ref{fig1}and Fig.\ \ref{fig2},   we have plotted the calculated RAS, ($Re\{ \frac{\Delta r} {r}\}$),  and SDA, 
  ($\epsilon_{[1\overline{1}0]}-\epsilon_{[001]}$), for Cu and Ag, respectively. 
To disentangle the effect of the intraband from interband part of $\epsilon^b$ in the RAS  we have also plotted  
in the upper panel of Fig.\ \ref{fig1} and  Fig.\ \ref{fig2}  
$Re \{\frac{\Delta r} {r}\} $ calculated without the intraband (Drude) term in $\epsilon^b$. 
 In Fig.\  \ref{fig3} we show the plots of  the calculated  real and imaginary part of $\epsilon^b$, separated into total and 
 interband parts, for bulk Cu (upper panel)  and Ag (lower panel). 
The  plots shown  in these  figures illuminate very  clearly that in both  
metals, the  overall  effect of $\epsilon^b$  is to re-shape profoundly the bare dielectric anisotropy
 of the surface  $\Delta\epsilon$,  depending on the energy and on the peculiarities of the bulk dielectric function of each metal.
 The RAS spectra of the two metals
may be divided into a low ($<$ 2.5 eV) and   high (between 3 and 5 eV) energy part, according to the prominent spectral features.
 The low energy peak  has a similar shape in both metals whereas the high energy structure  looks quite different.

Let us now focus on the two metals separately in order to highlight how characteristic quantities of the bulk, e.g. the plasma frequency,
affect the RAS spectrum.     
In  Fig.\ \ref{fig1} (upper panel) the RAS spectrum of Cu, with and without the Drude term in $\epsilon^b$  may be  compared with the   
 SDA in the lower panel. From this one sees very clearly      
that the low energy peak in the SDA undergoes a sign reversal and a little energy shift due to the intraband part of
$\epsilon^b$. This effect decreases with increasing energy until the interband part of $\epsilon^b$ takes over. 
For  energies  larger than $\simeq $5 eV the effect of the Drude tail dies off almost completely.
 The double peak shape of the high energy structure  is essentially determined by the  interband part of $\epsilon^b$  between 4 and 5
 eV (see Fig.\ \ref{fig3}, upper panel) whereas the Drude part of $\epsilon^b_1$  amplifies  the negative part of this peak 
between 4 and 4.5 eV.    

In Ag   the effect of $\epsilon^b$ is even more dramatic, as can be seen from the two curves  in the upper panel 
of Fig.\ \ref{fig2}, by  comparing  them  with the SDA plot in the lower panel. 
The low energy part of the spectrum (see the inset of Fig.\ \ref{fig2})
 is determined by the same interplay of bulk intraband $\epsilon^b$ and SDA as observed in Cu.
The RAS spectrum is, however,  dominated by the effect of the bulk plasmon producing the  overwhelming negative structure 
 at about 3.2 eV.  This  is 
due to the steep behaviour of $\epsilon^b$ close to the treshold of the onset of the interband transitions, as can be inferred from
Fig.\ \ref{fig3} (lower panel). Afterwards, the   spectrum is only sligtly affected by the Drude term, as in the case of Cu.

Let us now come to the direct comparison of our calculated  RAS spectra with the   corresponding experimental ones.
As pointed out very clearly by extensive studies\cite{hansen98,bremer99}, the lineshape of the observed structures may vary 
drastically not only
upon epxosure in air but also upon the number of cleaning cycles of the surface.

RAS spectra of Cu and Ag clean (1x1) (110) surface have been measured by various groups\cite{stahr98,hansen98,bremer99,stahr00,hofmann95}
 in different experimental conditions. All these spectra display the low energy peak at $\simeq$ 2.1 and 1.8, in Cu and Ag, respectively
  and the high energy structure  located between 4 and 5 eV, in agreement with our findings above. 
However,  the shape of the latter  structure depends upon treatment of the surface. 
In particular, a recent paper on the effect of sputtering on the RAS spectrum of the (110) surface of Cu shows that the shape 
of the double-peak structure at high energy (see Fig.\ \ref{fig1} ) varies very much with sputtering, holding as fixed the 'node' of 
the structure (see Fig.1 of Ref.\cite {bremer99}). This node occurs also in our calculations at essentially the same energy. 
 We emphasize that the position and overall shape of the two 
structures in the RAS of Cu presented in this letter agree quite satisfactorily with the experimental results\cite{hansen98,bremer99}.

Also for the Ag clean (1x1) (110) surface  our calculated RAS spectrum in Fig \ \ref{fig2} gives satisfactory agreement with the
 experimental
 findings apart from the shift of about 0.5 eV toward lower frequency in the location of the high energy structure. We attribute this
 discrepancy to the value of the threshold of the interband transitions that amounts to  3.3 eV in our calculations, as can be
 seen in Fig. \ \ref{fig3} (lower panel). This value is 0.5 eV smaller than the experimental one\cite{johnson&christy}. (This
 discrepancy has been found also in another calculation for bulk Ag, carried out using the pseudopotential method\cite{rubio}; 
 it must be due to  the  shortcoming of Density Functional Theory, that sometimes underestimates transition energies.)
 The lineshape of this structure is  surprisingly in agreement with the one obtained in very intensively polished samples, as
 discussed above\cite{hansen98}. 
 The double peak, resonance-like structure measured in freshly polished Ag surface, instead, does not occur
 in our {\it ab initio} calculations, since they do not include the long-ranged local-field effect. This is described   by the 
 'swiss-cheese'  phenomenological model\cite{mochan94} where  the   noble metal is modeled by a composite medium made by
   a jellium-like background  (the $sp$ electrons) screening the 
  embedded charged spheres, centred about ions  carrying  dipoles  (due to the $d$ electrons) interacting among themselves.
   The effect of the reduced symmetry of the surface on the dipole-dipole
 interaction leads to different plasma frequencies for the different light polarizations. This is the origin of the derivative-like 
 lineshape observed in Ag RAS at high energy. Its absence at surfaces treated by  
 several sputtering cycles  may be understood by the introduction of defects ({\it e.g.} vacancy islands) by the 
 treatment itself\cite{hansen98}. The geometrical roughening of the surface may   destroy the
 long range order, suppressing the effect of dipolar interactions and leaving only the dominant effect of the local environment, 
 well described by our one-electron calculations\cite{shkrebtii98}.
      
 To conclude and summarize, we have in this Letter  
  calculated {\it ab initio} the RAS spectra of Cu and Ag  (110) surfaces, showing  which structures are 
 due to the  surface dielectric anisotropy and which are due  to bulk effects. Moreover we have 
 disentagled intraband from interband contributions of the bulk dielectric function in  the RAS spectrum, showing that it is
 incorrect to interpret RAS spectra from the knowledge of the SDA only\cite{ecoss99-bouarab}. 
  For the first time the RAS structures observed at metal surfaces\cite{fernandez97,hansen98,bremer99} have been accurately reproduced 
 by {\it ab initio} calculations.   
  The measured lineshapes in the Cu and Ag RAS depend critically on 
 the conditions  of  the surface. As discussed exstensively in some papers, sputtering, annealing or exposure in air may change not
 only the amplitude but also the shape of the structures as found for the high-energy peaks\cite{hansen98,bremer99} in Cu and Ag.
 We conclude that our results and those obtained by the local-field model\cite{mochan94} refer to samples whose surface
  conditions\cite{hansen98,bremer99}   emphasize short-range and long-range arrangements of the atoms on the surface, respectively. 
\bigskip
 
*{\it On leave from }{Dipartimento di Fisica - Universit\'a dell'Aquila, I-67100 L'Aquila (Italy)}

\centerline{FIGURE CAPTIONS}
\bigskip

\begin{figure} \caption{RAS (upper panel, solid line) and SDA (lower panel) spectra for the  Cu-(110) surface. The dotted curve in the
upper panel is the RAS  obtained  excluding the Drude tail from $\epsilon^b$.} 

\label{fig1} 
\end{figure}

\begin{figure} \caption{The same plots as in Fig.\ \ref{fig1} for the  Ag-(110) surface. The inset shows the low energy peak on a
magnified scale} 

\label{fig2} 
\end{figure}

\begin{figure} \caption{Calculated Real and Imaginary parts of $\epsilon^b$ for Cu (upper panel) and Ag (lower panel): total
 (bold full line) and  interband (dotted line) $\epsilon^b_1$, total (dashed line) and interband (dash-dotted line) $\epsilon^b_2$.
  The bulk free-electron $\omega_p$  and relaxation time $\tau$ are
 taken from Ref. 11 to be 9.17 eV, 6.9 10$^{-15}$  $sec^{-1}$ and  9.22 eV, 31 10$^{-15}$   $sec^{-1}$ for Cu and Ag, respectively.} 

\label{fig3}
\end{figure}

\bigskip


\begin{references}

\bibitem{aspnes} {D.E.Aspnes, J. Vac. Sci. Technol. {\bf B3}, 1502 (1985)}

\bibitem{agras} {Y. Borensztein {\it et al.},  Phys. Rev. Lett. {\bf 71}, 2334 (1993)}

\bibitem{stahr98}{K. Stahrenberg {\it et al.},   Phys. Rev. {\bf B58}, R10207 (1998) }

\bibitem{fernandez97} V. Fernandez {\it et al.}, Surf. Sci. {\bf 377}, 388 (1997)

\bibitem{hansen98} J.K. Hansen, J. Bremer, O. Hunderi, phys. stat. sol. (a) {\bf 170}, 271 (1998)

\bibitem{bremer99} J. Bremer, J-K Hansen, O. Hunderi, Surf. Scie. Letters {\bf 436}, L735 (1999)
 
\bibitem{stahr00} {K. Stahrenberg {\it et al.}, Phys. Rev. {\bf B 61}, 3043 (2000) } 

\bibitem{cole98} R.J. Cole, B.G. Frederick, P. Weightman, J. Vac.Technol. {\bf A 15(5)}, 3088 (1998)

\bibitem{mochan94} W.L.Mochan {\it et al.}, {\bf 207A}, 334 (1994)


\bibitem{mrs99}P. Monachesi {\it et al.},  to appear in  {\bf Material Research Society}, Symposium Proceeding Series, vol. 579
 (MRS Fall Meeting 1999, Boston (USA))
 

\bibitem{ecoss99-bouarab}   M. Mebarki {\it et al.}, Proceedings of 'Ecoss99 Conference' to appear in Surf. Sci.
  
\bibitem{sda} J.D.E. McIntyre and D.E. Aspnes, Surf. Sci. {\bf 24}, 417 (1971)

\bibitem{hofmann95} {Ph. Hofmann {\it et al.},  Phys. Rev. Lett. {\bf 75} (1995)}

\bibitem{johnson&christy} { P.B. Johnson and R.W. Christy, Phys. Rev. {\bf B6}, 4370 (1972)}

\bibitem{wills} J. Wills and B.R. Cooper, Phys. Rev. {\bf B 36}, 3809 (1987)

\bibitem{andersen} O.K. Andersen, Phys. Rev. {\bf B 12}, 3060 (1975)

\bibitem{kohn-sham} W. Kohn and L.J. Sham, Phys. Rev. {\bf 140}, A1133 (1965) 

\bibitem{XC} L. Hedin and B.I. Lundquist, J. Phys. C {\bf 5}, 1629 (1972) 

\bibitem{tetrah} O. Jepsen and O.K. Andersen, Solid State Comm. {\bf 9}, 1763 (1971)

\bibitem{unrelx} Calculations performed with unrelaxed atomic positions give nothing but a small shift of the structures at
low energy  with respect to the present ones. 

\bibitem{relx} {M. Guillop\'e and B. Legrand, Surf. Sci. {\bf 215}, 577 (1989)}

\bibitem{rubio} M.A. Cazalilla {\it et al.} Phys. Rev. {\bf B 61}, 8033 (2000)

\bibitem{shkrebtii98} A.I. Shkrebtii {\it et al.}, Phys. Rev. Lett. {\bf 81}, 721 (1998) 
\end{references}
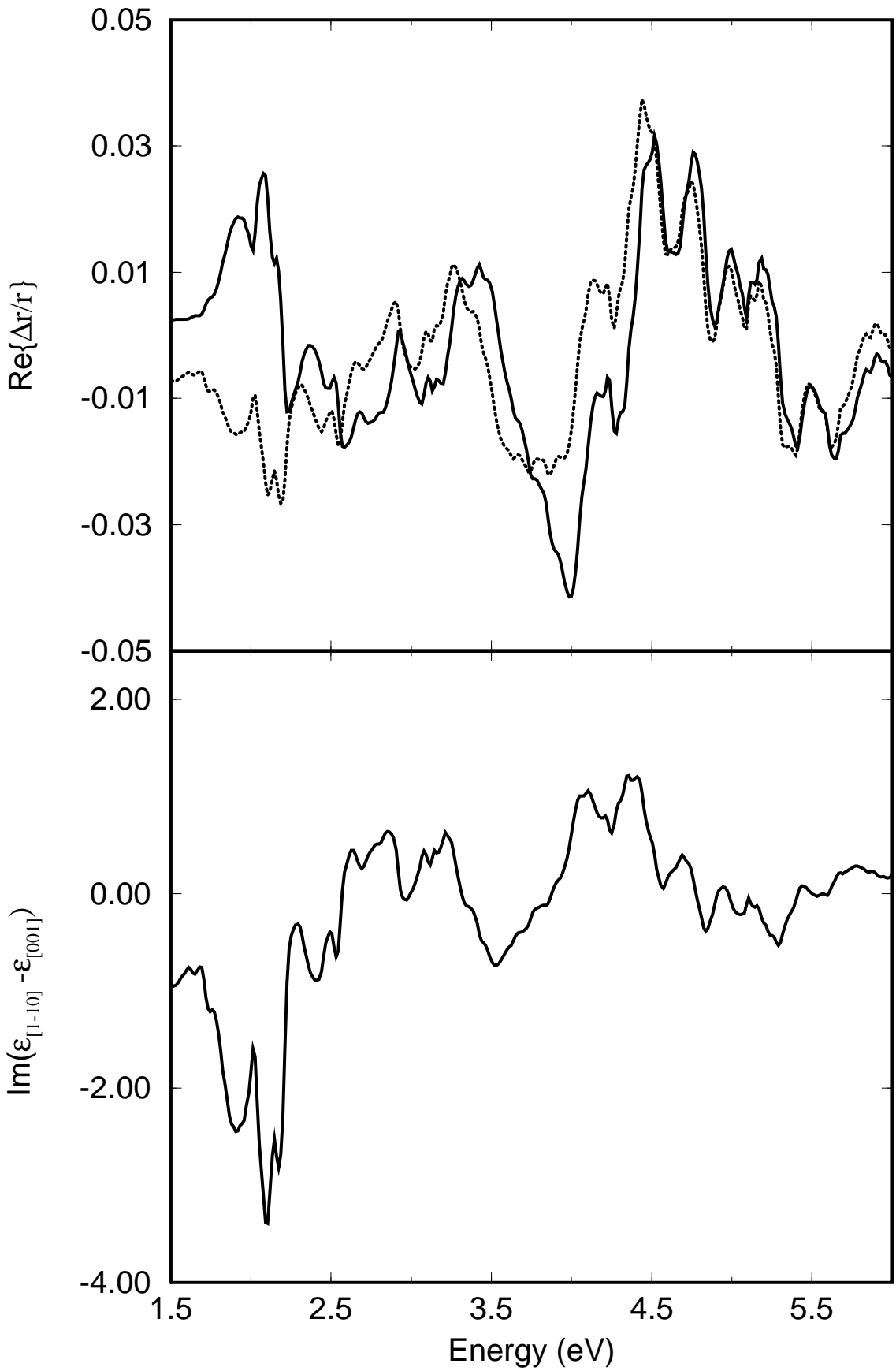
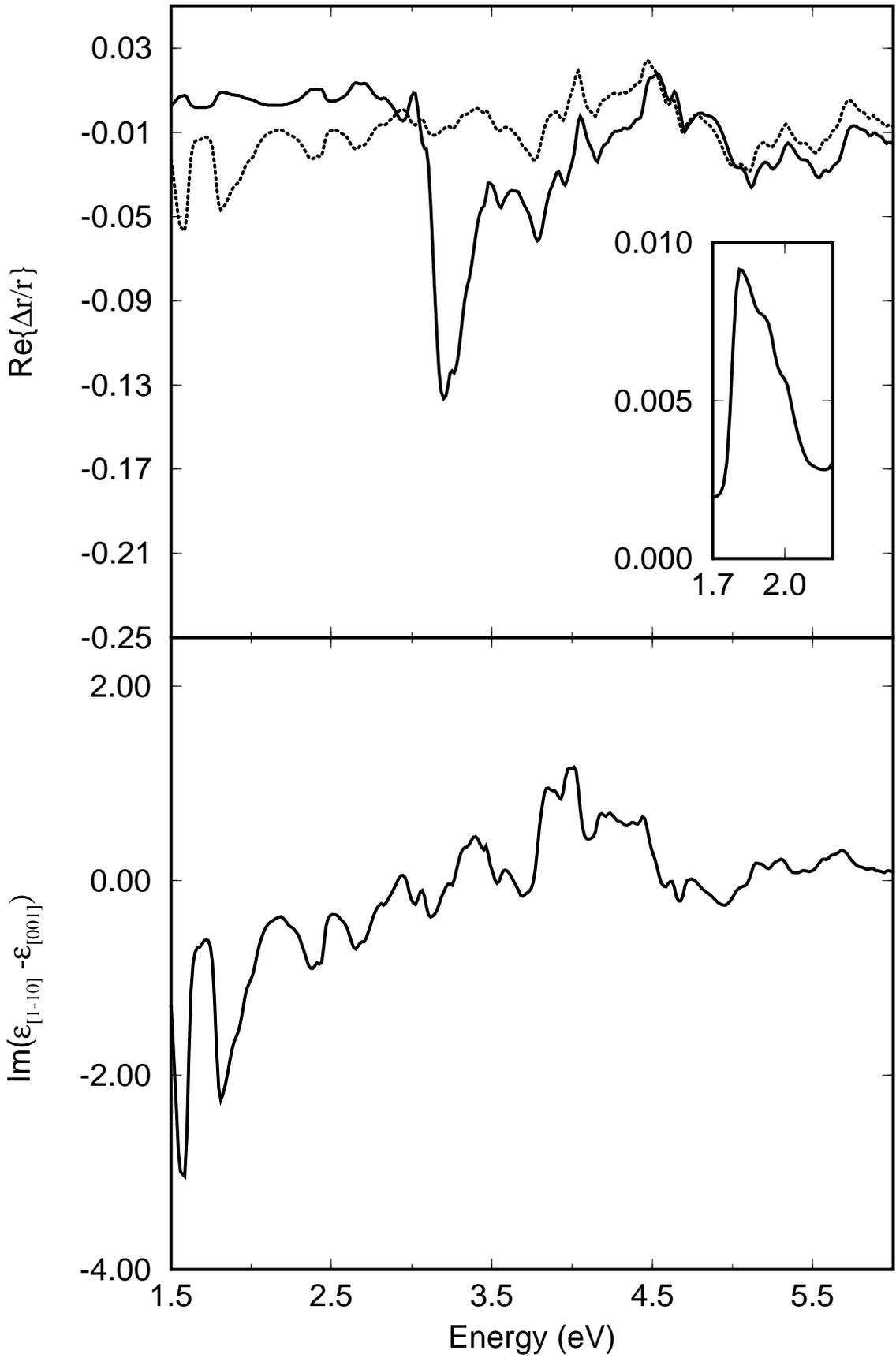
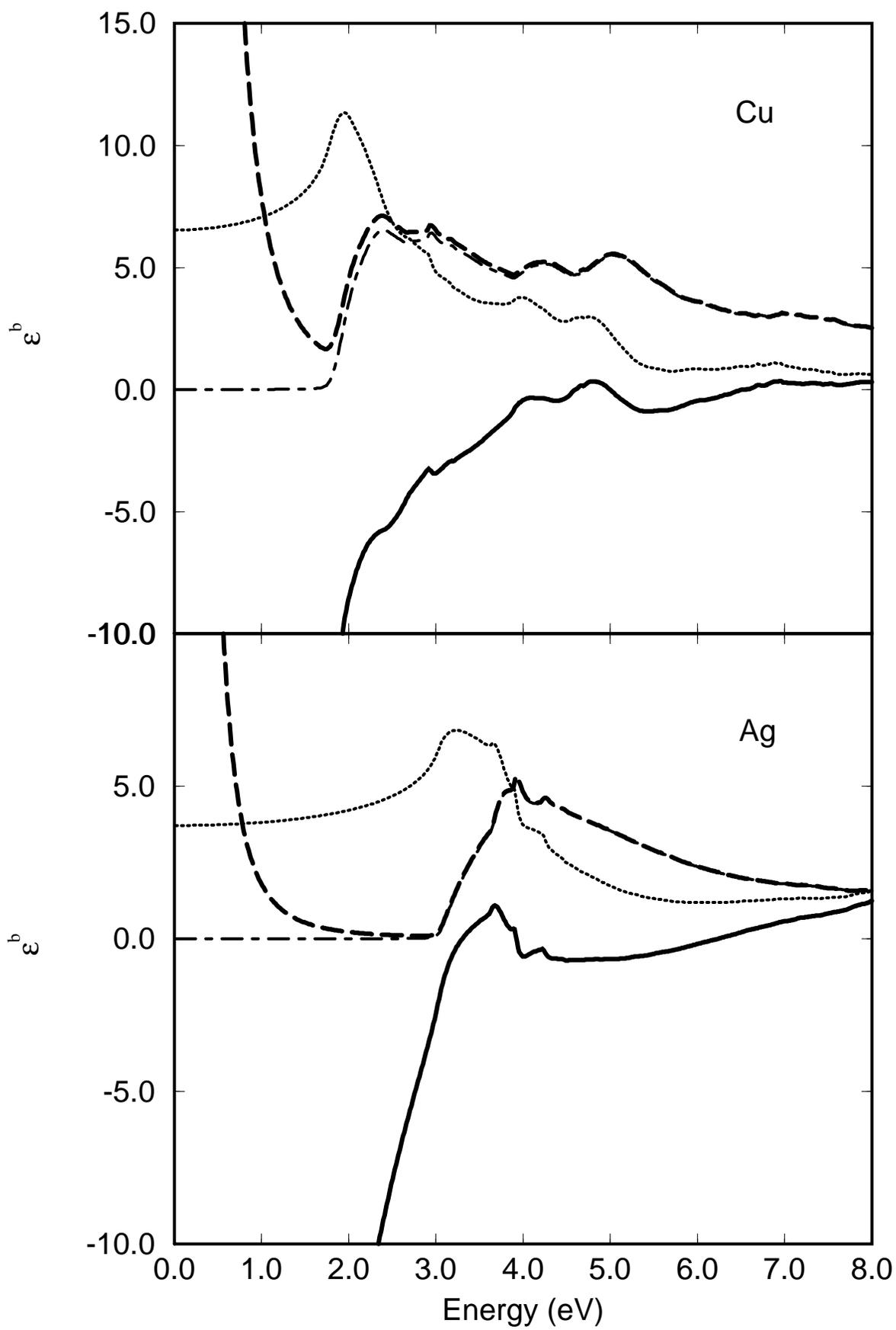
\end{document}